\RequirePackage{rotating}
\documentclass[fleqn,usenatbib]{mnras}

\usepackage{newtxtext,newtxmath}
\usepackage[T1]{fontenc}
\usepackage{ae,aecompl}
\usepackage{graphicx}
\usepackage{rotating}
\usepackage{lscape}
\usepackage{amsmath}	
\usepackage{amssymb}

\newcommand{\mr}{\mathrm}

\usepackage{mathrsfs}

\usepackage{bm}
\usepackage{url}
\graphicspath{{figs/}}
\newcommand{\comm}[1]{}
\usepackage{enumerate}
\usepackage{caption}
\usepackage{tikz}

\usepackage{stackengine}

\usepackage{threeparttable}
\usepackage{tabulary}
\newcolumntype{K}[1]{>{\centering\arraybackslash}p{#1}}
\usepackage{cellspace}
\DeclareCaptionType{diag}[Table]

\title[H$_2$O abundances in 10 hot giant planets]{H$_2$O abundances and cloud properties in ten hot giant exoplanets}
\author[Pinhas et al.]{
Arazi Pinhas,$^{1}$\thanks{E-mail: ap817@ast.cam.ac.uk}
Nikku Madhusudhan,$^{1}$\thanks{E-mail: nmadhu@ast.cam.ac.uk}
Siddharth Gandhi,$^{1}$
Ryan J. MacDonald$^{1}$
\\
$^{1}$Institute of Astronomy, University of Cambridge, Madingley Road, Cambridge, CB3 0HA, UK
}

\date{}
\pubyear{2018}

\begin{document}
\label{firstpage}
\pagerange{\pageref{firstpage}--\pageref{lastpage}}
\maketitle

\begin{abstract}
Transmission spectroscopy of exoplanets has the potential to provide precise measurements of atmospheric chemical abundances, in particular of hot Jupiters whose large sizes and high temperatures make them conducive to such observations. To date, several transmission spectra of hot Jupiters have revealed low amplitude features of water vapour compared to expectations from cloud-free atmospheres of solar metallicity. The low spectral amplitudes in such atmospheres could either be due to the presence of aerosols that obscure part of the atmosphere or due to inherently low abundances of H$_2$O in the atmospheres. A recent survey of transmission spectra of ten hot Jupiters used empirical metrics to suggest atmospheres with a range of cloud/haze properties but with no evidence for H$_2$O depletion. Here, we conduct a detailed and homogeneous atmospheric retrieval analysis of the entire sample and report the H$_2$O abundances, cloud properties, terminator temperature profiles, and detection significances of the chemical species. Our present study finds that the majority of hot Jupiters have atmospheres consistent with sub-solar H$_2$O abundances at their day-night terminators. The best constrained abundances range from $\mr{log(H_2O)}$ of $-5.04^{+0.46}_{-0.30}$ to $-3.16^{+0.66}_{-0.69}$, which compared to expectations from solar-abundance equilibrium chemistry correspond to $0.018^{+0.035}_{-0.009}\times$ solar to $1.40^{+4.97}_{-1.11}\times$ solar. Besides H$_2$O we report statistical constraints on other chemical species and cloud/haze properties, including cloud/haze coverage fractions which range from $0.18^{+0.26}_{-0.12}$ to $0.76^{+0.13}_{-0.15}$. The retrieved H$_2$O abundances suggest sub-solar oxygen and/or super-solar C/O ratios, and can provide important constraints on the formation and migration pathways of hot giant exoplanets.
\end{abstract}

\section{Introduction}\label{intro}

A recent spectroscopic survey investigated the atmospheric properties of ten hot Jupiters using transmission spectra in the 0.3 - 5.0 $\mu$m range  \citep{sing16}. The planets, all similar in size to Jupiter, range in equilibrium temperature between 900 K and 2600 K and in mass between 0.2 and 1.5 Jupiter masses. The spectra for the objects were obtained with multiple instruments including the STIS (0.3 - 1.0 $\mu$m) and WFC3 (0.8 - 1.8 $\mu$m) spectrographs aboard the Hubble Space Telescope (HST) and the Spitzer IRAC photometric channels at 3.6 $\mu$m and 4.5 $\mu$m. The amplitudes of H$_2$O features across the sample were found to be low, below 2-3 atmospheric scale heights. Such low amplitudes were interpreted in the past as either due to obscuration by clouds/hazes \citep[e.g.,][]{deming13} or due to inherently low H$_2$O abundances \citep[e.g.,][]{madhu14b}

\citet{sing16} interpreted the observed atmospheric spectra using empirical metrics based on chemical equilibrium atmospheric models to constrain the prominence of clouds/hazes vis-\`a-vis the H$_2$O abundances in the atmospheres. The contribution of H$_2$O was represented by the amplitude of the H$_2$O feature at $\sim$1.4 $\mu$m, while the difference between the optical and infrared planetary radii was taken as representative of the cloud/haze contribution. These empirical metrics of the observations were evaluated with a grid of theoretical chemical equilibrium models to suggest that the atmospheres spanned a continuum from clear to cloudy with no evidence for sub-solar H$_2$O abundances. However, a complementary study of the original \citet{sing16} survey by \cite{barstow17} suggested a general range of H$_2$O abundances indicating sub-solar abundances in most of the targets.

In order to investigate the atmospheres in detail it is imperative to quantitatively infer the H$_2$O abundances and cloud/haze contributions simultaneously for all the planets in the sample with the fewest possible model assumptions and in a statistically robust manner. Such abundance estimates of individual objects as well as populations with the derived statistical uncertainties would be valuable inputs for studies investigating planet formation and migration \citep{oberg11,mousis12, madhu14a,mordasini16,madhu17}. State-of-the-art Bayesian atmospheric retrieval techniques make it possible to estimate such atmospheric properties from transmission spectra \citep{madhu14b,line13,kreidberg14b,macdonald17_209}.

\citet{barstow17} carried out a study of the ten hot Jupiters \citep{sing16} using the Non-linear Optimal Estimator for MultivariatE Spectral analySIS (NEMESIS) algorithm to infer properties of their atmospheres. They found that all spectra are consistent with the presence of clouds and hazes. However, contrary to the work of \citet{sing16}, \citet{barstow17} also report that all their planets have sub-solar H$_2$O abundances between 0.01$\times$ solar and solar. The optimal estimation retrieval technique used in \citet{barstow17} assumes Gaussian priors around a single best-fit solution and doesn't allow a full marginalisation of parameter distributions such that recovering statistical constraints of model parameters and their significances is not possible.

In the present study we conduct a homogeneous Bayesian retrieval analysis of the ten hot giant exoplanets contained in \citet{sing16} to determine statistical estimates of their atmospheric properties. The estimated atmospheric properties include the H$_2$O and other chemical abundances, the cloud/haze properties, and the temperature profiles. In addition to parameter estimation of the atmospheric model, the statistical sampling method used allows us to extract the full marginalisation of the likelihood function, the evidence $\mathcal{Z}$. Using $\mathcal{Z}$ allows a comparison of models with different input physics and a measurement of detection significances of various parameters, notably the H$_2$O abundance. By this analysis we obtain best-fit inferences and statistically significant Bayesian credible intervals for the H$_2$O abundances and other atmospheric parameters.

The paper is organised as follows. In Section \ref{obs} we outline the atmospheric transmission observations that are used. We then shortly outline our atmospheric retrieval framework in Section \ref{retrieval framework}. In Section \ref{results} we present results from our retrieval analysis, with a focus on H$_2$O abundances. We then discuss our work in light of previous studies \citep{sing16,barstow17} and review the essential outcomes of our study in Section \ref{conclusions}.

\section{Observations}\label{obs}

We use a retrieval approach to interpret transmission observations of 10 hot Jupiters contained in \citet{sing16}. Table \ref{table:planets} shows the properties of each planetary system along with details of the observations used as input to our retrievals. The transmission spectra used in our work are obtained from recent studies coverings a broad wavelength range from 0.3-5.0 $\mu$m using the {\it HST} and {\it Spitzer} facilities \citep{sing16,kreidberg15,line13, tsiaras17}. In particular, the data are products of eight observing modes: HST STIS G430L, HST STIS G750L/M, HST ACS G800L, HST WFC3 G102, HST WFC3 G141, and Spitzer IRAC photometry bandpasses at $3.6\,\mu$m and $4.5\,\mu$m. All planets except for WASP-6b have data comprising at least HST STIS G430L, HST STIS G750L/M, HST WFC3 G141, and the two Spitzer IRAC channels. WASP-6b lacks WFC3 spectroscopy while HD 189733b has additional spectroscopy in the 0.8-1.1 $\mu$m range from HST ACS G800L.

We use the same data as in a recent spectral survey study \citep{sing16} except in the cases of HAT-P-12b, WASP-12b, and WASP-39b where other and/or additional data have been used in the near infrared \citep{kreidberg15,line13, tsiaras17}. In addition, we have not used the NICMOS data for HD 189733b as it has been shown that its systematics cannot be reliably understood and corrected at the level needed to detect molecular absorption in hot Jupiters \citep{gibson11}. In the case of HAT-P-12b, we use HST WFC3 G141 spectra with 23 data points \citep{line13} as compared with 11 G141 data points \citep{sing16}. Different G141 and additional G102 observations were used for WASP-12b \citep{kreidberg15}. This was done for two reasons. First, the \citet{kreidberg15} WASP-12b G141 data constitute a combination of six transits compared with one transit for the spectral survey study of \citet{sing16}. The \citet{kreidberg15} data were also obtained in spatial scanning mode as opposed to staring mode, allowing for higher-precision transit depths. Both the increased transit count and spatial-scan observing mode result in a median precision of 51 ppm on the used data, while the best precision achieved by \citet{sing16} is 100 ppm. Retrieving the higher-precision data translates into tighter constraints on WASP-12b's average terminator H$_2$O abundance. Second, additional data with the G102 grism are also used and have precisions similar to the G141 grism spectroscopy \citep{kreidberg15}. The additional information content contained in the G102 data between 0.78 and 1.07 $\mu$m provides for generally tighter constraints on the retrieved parameters. In the case of WASP-39b, additional WFC3 G141 data \citep{tsiaras17} are used to compensate for lack of near-infrared data in the survey study \citep{sing16}.

The average precisions on the observed spectra span a wide range from high precisions of $\sim$30 ppm (HD 209458b G141 data) to low values of $\sim$400 ppm (WASP-12b Spitzer photometry). This suite of observations constitutes a baseline sample to study with retrieval techniques since the datasets have undergone a consonant reduction process for systematics \citep{sing16}, with the exception of the HAT-P-12b G141 data \citep{line13}, WASP-12b G102 and G141 data \citep{kreidberg15}, and WASP-39b G141 data \citep{tsiaras17} included in our work.

\begin{table*}
\centering 
\begin{tabular}{c c c c c c c}
\hline
\hline
\rule{0pt}{3ex}  Planet & $T_{\mr{eq}}$ (K) & $R_{p}$ ($R_J$) & $M_p$ ($M_J$) & log ($g(\mr{cms^{-2}}$)) & $R_{\star}$ ($R_{\odot}$) & Instrument/Sub-Instrument/Disperser or Detector\\ [1ex]
\hline
\\ [-1.5ex]
\rule{0pt}{3ex}
 HAT-P-12b & 960 & 0.96 & 0.21 & 2.439 & 0.720 & \parbox{5cm}{ \centering HST/STIS/G430L\\ \centering HST/STIS/G750L \\ \centering HST/WFC3 IR/G141$^{\blacklozenge}$\\ \centering Spitzer/IRAC/3.6 channel\\ \centering  Spitzer/IRAC/4.5 channel}\\[1ex]
 
 \\
 
 WASP-39b & 1,120 & 1.27 & 0.28 & 2.613 & 0.910 & \parbox{5cm}{ \centering HST/STIS/G430L\\ \centering HST/STIS/G750L \\ \centering HST/WFC3 IR/G141$^{\blacklozenge}$\\ \centering Spitzer/IRAC/3.6 channel\\ \centering  Spitzer/IRAC/4.5 channel}\\[1ex]
 
 \\
 
 WASP-6b & 1,150 & 1.22 & 0.50 & 2.940 & 0.864 & \parbox{5cm}{ \centering HST/STIS/G430L\\ \centering HST/STIS/G750L \\ \centering Spitzer/IRAC/3.6 channel\\ \centering  Spitzer/IRAC/4.5 channel}\\[1ex]
 
 \\
 
 HD 189733b & 1,200 & 1.14 & 1.14 & 3.330 & 0.751 & \parbox{5cm}{ \centering HST/STIS/G430L\\ \centering HST/STIS/G750M \\ \centering HST/ACS/G800L \\ \centering HST/WFC3 IR/G141 \\ \centering Spitzer/IRAC/3.6 channel\\ \centering  Spitzer/IRAC/4.5 channel}\\[1ex]
 
 \\
 
 HAT-P-1b & 1,320 & 1.32 & 0.53 & 2.875 & 1.150 & \parbox{5cm}{ \centering HST/STIS/G430L\\ \centering HST/STIS/G750L \\ \centering HST/WFC3 IR/G141 \\ \centering Spitzer/IRAC/3.6 channel\\ \centering  Spitzer/IRAC/4.5 channel}\\[1ex]
 
 \\
 
 HD 209458b & 1,450 & 1.359 & 0.69 & 2.9634 & 1.155 & \parbox{5cm}{ \centering HST/STIS/G430L\\ \centering HST/STIS/G750L \\ \centering HST/WFC3 IR/G141 \\ \centering Spitzer/IRAC/3.6 channel\\ \centering  Spitzer/IRAC/4.5 channel}\\[1ex]
 
 \\
 
 WASP-31b & 1,580 & 1.55 & 0.48 & 2.663 & 1.274 & \parbox{5cm}{ \centering HST/STIS/G430L\\ \centering HST/STIS/G750L \\ \centering HST/WFC3 IR/G141 \\ \centering Spitzer/IRAC/3.6 channel\\ \centering  Spitzer/IRAC/4.5 channel}\\[1ex]
 
 \\
 
 WASP-17b & 1,740 & 1.89 & 0.51 & 2.556 & 1.578 & \parbox{5cm}{ \centering HST/STIS/G430L\\ \centering HST/STIS/G750L \\ \centering HST/WFC3 IR/G141 \\ \centering Spitzer/IRAC/3.6 channel\\ \centering  Spitzer/IRAC/4.5 channel}\\[1ex]
 
 \\
 
 WASP-19b & 2,050 & 1.41 & 1.14 & 3.152 & 1.027 & \parbox{5cm}{ \centering HST/STIS/G430L\\ \centering HST/STIS/G750L \\ \centering HST/WFC3 IR/G141 \\ \centering Spitzer/IRAC/3.6 channel\\ \centering  Spitzer/IRAC/4.5 channel}\\[1ex]
 
 \\
 
 WASP-12b & 2,510 & 1.73 & 1.40 & 3.064 & 1.506 & \parbox{5cm}{ \centering HST/STIS/G430L\\ \centering HST/STIS/G750L \\ \centering HST/WFC3 IR/G102$^{\blacklozenge}$\\ \centering HST/WFC3 IR/G141$^{\blacklozenge}$\\ \centering Spitzer/IRAC/3.6 channel\\ \centering  Spitzer/IRAC/4.5 channel}\\[1ex]
 \\[-1.5ex]
\hline
\hline
\end{tabular}
\parbox{15.5cm}{\caption{Planetary system properties and observations. All data are from the spectral survey \citep{sing16} except for WFC3 IR data for HAT-P-12b \citep{line13}, WASP-12b \citep{kreidberg15}, and WASP-39b \citep{tsiaras17}, the latter three marked by $^{\blacklozenge}$. The host star radii were calculated through $R_{p}/R_{\star}$ ratios from various discovery papers: HAT-P-12b \citep{line13}, WASP-39b \citep{fischer16}, WASP-6b \citep{nikolov15}, HD 189733b \citep{mccullough14}, HAT-P-1b \citep{nikolov14}, HD 209458b \citep{deming13}, WASP-31b \citep{sing15}, WASP-17b \citep{mandell13}, WASP-19b \citep{huitson13}, and WASP-12b \citep{sing13}. $T_{\mr{eq}}$, $R_p$, $M_p$, and log($g$) are from the spectral survey \citep{sing16}.}
\label{table:planets}}
\end{table*}

\section{Atmospheric Retrieval Method}\label{retrieval framework}

We use a Bayesian retrieval method to infer the atmospheric properties along the planetary terminator of the hot Jupiters from their observed transmission spectra. We use an atmospheric retrieval code for transmission spectra, {\sc Aura} \citep{pinhas18_Aura}, that is adapted from \citet{gandhi18} and follows methods in \citet{madhu09} and \citet{macdonald17_209} as discussed below. Retrieving model parameters of transmission spectra requires two basic components: a forward model and a statistical sampling algorithm. We here outline these in turn.

\subsection{Forward Model}

First, the pressure-temperature ($p$--$T$) profile in the atmosphere follows a one-dimensional prescription that captures the terminator-averaged temperature with height \citep{madhu09}. The parametric $p$--$T$ profile is able to fit disparate planetary atmosphere structures, mimicking atmospheric conditions of solar system planets as well as exoplanetary atmosphere models in the literature \citep{madhu09}. The atmosphere is sectioned into three zones with the following equations for the temperature, 
\begin{align}
    T &= T_0 + \left ( \frac{\mr{ln}(P/P_0)}{\alpha_1} \right )^2 \hspace{0.5cm} P_0 < P < P_1,  \\
    T &= T_2 + \left ( \frac{\mr{ln}(P/P_2)}{\alpha_2} \right )^2 \hspace{0.5cm} P_1 < P < P_3,  \\
    T &= T_2 + \left ( \frac{\mr{ln}(P_3/P_2)}{\alpha_2} \right )^2 \hspace{0.5cm} P > P_3. 
\end{align}
In total, the profile consists of six free parameters: $T_0$, $\alpha_1$, $\alpha_2$, $P_1$, $P_2$, and $P_3$. The temperature at the top of the atmosphere is $T_0$; $\alpha_1$ and $\alpha_2$ are variables responsible for the gradient of the profile; and $P_1$, $P_2$, and $P_3$ are pressure values that define the three distinct zones. In the primary transit case, for which there is no source function and thus no potential thermal inversion, $P_2 \leq \, P_1 < P_3$. We partition our model atmosphere into 100 layers spaced equally in log-pressure between $10^{-6}$ bar and $10^{2}$ bar. While the lower level in the atmosphere of 100 bar is well below the observable photosphere, the upper level of $10^{-6}$ bar marks the region where molecules are  photodissociated \citep{moses11,moses13,moses14} and hence no longer contribute opacity. In addition, a seventh free parameter is the reference pressure $P_{\mr{ref}}$, the a priori unknown pressure at $R_p$.

In addition to the temperature structure of the atmosphere, the retrieval forward model contains a suite of chemical species. The average terminator volume mixing ratios $X_{\mr{i}} =  n_{\mr{i}}/n_{\mr{tot}}$ of chemical species are free parameters of the model. We consider important chemical species having significant spectral features from 0.3-5.0 $\mu$m. These include H$_2$O, CH$_4$, NH$_3$, HCN, CO, CO$_2$, Na and K in addition to the main H$_2$ and He constituents in gaseous planets. Our model also includes collisionally-induced-absorption (CIA) opacities for H$_2$-H$_2$ and H$_2$-He as well as H$_2$ Rayleigh scattering. Molecular and collisionally-induced cross sections are computed by \citet{gandhi17} from various line-list databases \citep{rothman10, rothman13, richard12, tennyson16}. In particular, our CH$_4$, HCN, and NH$_3$ molecular line data are from EXOMOL \citep{tennyson16, yurchenko11, barber14, yurchenko14} and the line data for H$_2$O, CO, and CO$_2$ are obtained from HITEMP \citep{rothman10}. Some partition functions used to calculate the molecular line strengths are limited in temperature. In such cases a cubic extrapolation is performed. The CIA data are sourced from the HITRAN archive \citep{richard12}.

Finally, our forward model includes a parameterised model for in-homogeneous clouds and hazes, accounting for effects due to a range of particle sizes \citep{macdonald17_209}.
These terms are typically used in reference to the predominant morphology of spectral features they induce in transmission spectra, especially in parametric models used for atmospheric retrieval or forward models \citep[e.g.,][]{benneke12, kreidberg14b, sing16, macdonald17_209}. A `haze' is represented by a non-grey opacity in the optical wavelength region and through a power-law dependence on wavelength, while a `cloud' is generally used to mean a source of grey opacity from particle sizes $\gtrsim$1 $\mu$m throughout the spectral range effective to a certain height in the atmosphere. Therefore, hazes produce Rayleigh-like slopes in the visible spectral range while clouds produce grey opacity throughout the spectrum \citep{pinhas17}. The extinction coefficient (with units of inverse length) which broadly incorporates these two spectral effects is, 
\begin{equation}
    \kappa_{\mr{cloud/haze}} (r) = \left\{\begin{array}{ll}
    n_{\mr{H_2}} a \sigma_0(\lambda/\lambda_0)^{\gamma}\,\,\,\,\,\,\,\,\,\,{ \it if}\,\,P<P_{\mr{cloud}}\\
                  \infty\,\,\,\,\,\,\,\,\,\,\,\,\,\,\,\,\,\,\,\,\,\,\,\,\,\,\,\,\,\,\,\,\,\,\,\,\,\,\,\,\,\,\,\,\,\mr{otherwise}\\
                \end{array}
              \right.
\end{equation}
where the first relation represents a slope in the optical characteristic of hazes and the second equality represents a grey opacity across all wavelengths characteristic of clouds of particle sizes $\gtrsim$1 $\mu$m. Here, $\lambda_0$ is a reference wavelength (0.35 $\mu$m), $\sigma_0$ is the gaseous H$_2$ Rayleigh scattering cross-section at $\lambda_0$ ($5.31\times 10^{-31}\mr{m^2}$), $a$ is the `Rayleigh-enhancement factor' and $\gamma$ is the `scattering slope'. The Rayleigh-enhancement factor effectively quantifies the offset level of the optical transmission spectrum. In principle, inherent variations in species' refractive indices can constrain different species through the value of $a$. Moreover, the optical slope can also be used to constrain species since each one has characteristic values of $\gamma$ \citep{pinhas17}. The enhancement factor is related to the haze mixing ratio through $X_{\mr{haze}} = X_{\mr{H_2}} a \frac{\sigma_0}{\sigma_{\mr{haze,0}}}$, in which $\sigma_{\mr{haze,0}}$ is the haze cross-section at $\lambda_0$. The $\sigma_{\mr{haze,0}}$ value is obtained from a recent study \citep{pinhas17} that uses experimental refractive index data for a dozen condensates and shows their extinction cross-sections for small particles of $\sim$$10^{-2}\mu$m at $\lambda_0$ to be $\sim$$10^{-12}\,{\mr{cm}}^{2}$. Three cloud/haze parameters for our retrieval are thus $a$, $\gamma$, and $P_{\mr{cloud}}$.

The fourth cloud/haze parameter $\bar{\phi}$ describes the terminator-averaged cloud/haze contribution of a two-dimensional planetary atmosphere and enters into the measured or effective transit depth as
\begin{equation}\label{effective_transit_depth}
    \Delta_{\mr{planet}}(\lambda) = \bar{\phi} \Delta_{\mr{cloud/haze}} (\lambda)  + (1 - \bar{\phi}) \Delta_{\mr{clear}}(\lambda).
\end{equation}
where $\Delta_{\mr{cloud/haze}}$ ($\Delta_{\mr{clear}}$) are the transit depths computed with (without) incorporation of clouds/hazes. The transit depth components in the equation above are calculated through equation (A10) in \citet{macdonald17_209}. A terminator completely covered with clouds/hazes has $\bar{\phi} = 1$ while a clear atmosphere along the terminator has $\bar{\phi} = 0$. A terminator with $0<\bar{\phi}<1$ contains patchy or in-homogeneous clouds and hazes. In totality, our retrieval forward model contains a maximum of 19 free parameters: $T_0$, $\alpha_1$, $\alpha_2$, $P_1$, $P_2$, $P_3$, $P_{\mr{ref}}$, $X_{\mr{Na}}$, $X_{\mr{K}}$, $X_{\mr{H_2O}}$, $X_{\mr{CH_4}}$, $X_{\mr{NH_3}}$, $X_{\mr{HCN}}$, $X_{\mr{CO}}$, $X_{\mr{CO_2}}$, $a$, $\gamma$, $P_{\mr{cloud}}$, and $\bar{\phi}$.

\subsection{Statistical Module}

A statistical parameter estimation algorithm is used to retrieve the atmospheric parameters of the forward model given a set of data. The statistical algorithm uses the observations to find commensurate posterior distributions of forward model parameters and their credibility intervals. Here we only briefly emphasize the utility of our statistical approach, while a detailed discussion is available in several studies \citep{skilling06, macdonald17_209,pinhas18_Aura}. The statistical framework uses the {\sc MultiNest} nested sampling technique that enables model parameter estimation and calculation of the Bayesian evidence \citep{skilling04, feroz08, feroz09, feroz13}. {\sc MultiNest} is implemented through a python wrapper, PyMultiNest \citep{buchner14}. The multi-dimensional parameter space is explored with 4,000 live points, a middle-way in maximizing the accuracy of the computed evidence ($\mathcal{Z}$) and minimizing the total time to reach a converged solution. In addition to model parameter estimation, the statistical retrieval approach of \citet{skilling04} allows full marginalisation of the likelihood function to compute the evidence $\mathcal{Z}$. The $\mathcal{Z}$ statistic enables model comparison for different model scenarios (e.g., clear versus cloudy) and calculation of detection significances for various chemical species.

\section{Results}\label{results}

Before applying our retrieval method to observations we performed synthetic retrievals for each planet to test the fidelity of our retrieval framework. We generated synthetic data for a chosen set of 19 forward model parameters. The precisions of the synthetic data were chosen assuming median values on the precisions associated with the actual data from each instrument. For the majority of planets, this translated into different precisions for five instruments: HST STIS G430L, HST STIS G750L, HST WFC3 G141, and two Spitzer bandpasses at 3.6 and 4.5 $\mu$m. In addition to using representative uncertainties, the simulated data were shifted with random Gaussian noise drawn from distributions with standard deviations matching the precisions in order to resemble genuine observations. The majority of parameter values used to generate the synthetic data for each planet were retrieved within the 1$\sigma$ intervals.

We show examples of simulated retrievals for data quality representative of WASP-12b, which has relatively moderate data quality compared to the planets in the ensemble. Simulated retrievals were conducted for a range of atmospheric $p$--$T$ profiles and abundances and the results are available on the Open Science Framework\footnote{\label{OSFlink}\url{https://osf.io/wtqjr/?view_only=8f70b80b5d834f3d8ef8be4ee7b77b27}}. These simulations illustrate the self-consistency of the retrieval method. In the case of WASP-19b and WASP-6b, the number of observations is less than the number of model parameters and the retrieved parameter posteriors are somewhat more broad. Nevertheless, a simulated retrieval for data quality of WASP-19b demonstrates that the retrieved posteriors do not show distributions significantly shifted from the true values which would be diagnostic of fitting to the observational noise. The retrieval forward model has also been validated against results from radiative transfer codes of other groups \citep{fortney10,deming13,line13,heng17}, with which we find good agreement.

\begin{table}
\centering
\parbox{6.5cm}{
\caption{Prior information used in the retrieval analyses.}
\label{tab:ret_priors}} 
\begin{tabular}{K{1.4cm}  K{1.8cm} K{3.2cm}}
\hline
\hline
\rule{0pt}{3ex}  
Parameter & Prior Distribution & Prior Range\\ [1ex]
\hline
\rule{0pt}{3ex}
\\[-3ex]
$T_0$ & Uniform & \parbox{3cm}{ \centering 400 $-$ $T_{\mr{eq}}+200$ K (for $T_{\mr{eq}} < 1200$ K) \\ \centering 800 $-$ $T_{\mr{eq}}+200$ K (for $T_{\mr{eq}} > 1200$ K)} \\[1ex]
$\alpha_{1,2}$ & Uniform & $0.02 - 1$ K$^{-1/2}$ \\
$P_{1,2}$ & Log-uniform & $10^{-6} - 10^{2}$ bar\\
$P_{3}$ & Log-uniform & $10^{-2} - 10^{2}$ bar\\
$X_i$ & Log-uniform & $10^{-12} - 10^{-2}$ \\
$a$ & Log-uniform & $10^{-4} - 10^{8}$ \\
$\gamma$ & Uniform & -$20 - 2$ \\
$P_{\mr{cloud}}$ & Log-uniform & $10^{-6} - 10^{2}$ bar\\
$\bar{\phi}$ & Uniform & $0 - 1$ \\
\hline
\hline
\end{tabular}
\end{table}

\begin{figure*}
    \centering
    \includegraphics[scale=1.01]{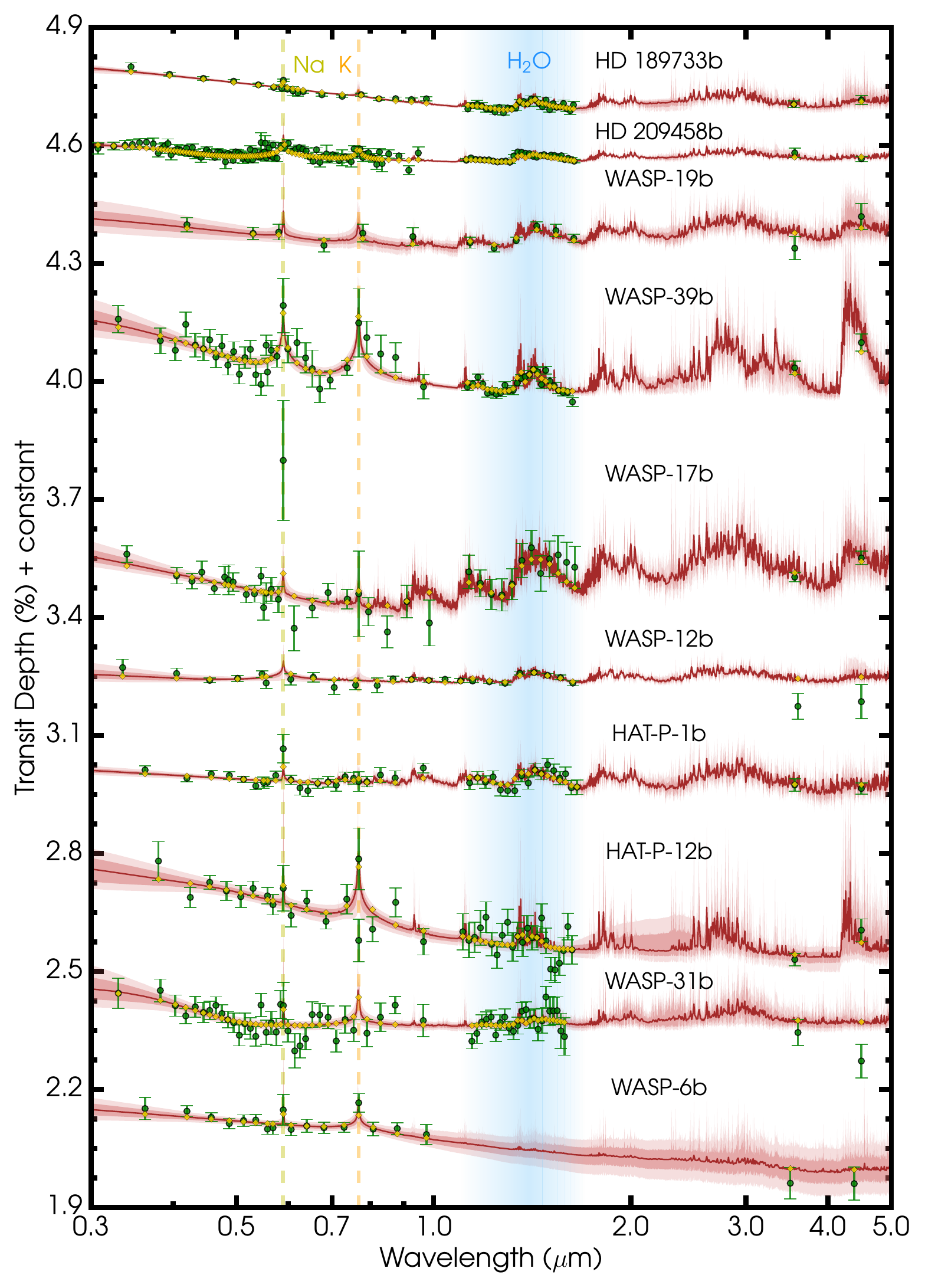}
\caption{Retrieved model transmission spectra compared to observations for the ensemble of hot Jupiters. The data are shown in green and the retrieved median model is in dark red with associated 1$\sigma$ and 2$\sigma$ confidence contours. The yellow diamonds are the binned median model at the same resolution as the observations. The best-fit median model in dark red has been smoothed for clarity. The data are discussed in Section \ref{obs} and shown in Table \ref{table:planets}.}
\label{fig:TD_bestfits}
\end{figure*}

In the present study we conduct a homogeneous Bayesian retrieval analysis on observations of ten hot giant planets (see Section \ref{obs}) to determine statistical estimates of their atmospheric properties. The ensemble of hot Jupiters includes HAT-P-12b, WASP-39b, WASP-6b, HD 189733b, HAT-P-1b, HD 209458b, WASP-31b, WASP-17b, WASP-19b, and WASP-12b. The prior distributions and ranges of the retrievals are shown in Table \ref{tab:ret_priors}. The estimated atmospheric properties include the H$_2$O and other chemical abundances, cloud/haze properties, and temperature profiles, along with detection significances for the chemical species. Each planet was sampled with about 5 million models, for a total of more than 60 million model runs. Using our framework, we derive marginalised posterior probability distributions and statistical estimates for each atmospheric parameter. A panorama of the retrieved model fits to the observations are shown in Figure \ref{fig:TD_bestfits}, while the full set of our retrieval results including the posterior distributions, best-fit spectra, and $p$--$T$ profiles are available on the Open Science Framework\textsuperscript{\ref{OSFlink}} and in Tables \ref{tab:retvals1}-\ref{tab:retvals3}. The properties of each planetary system along with details of observations used as input to our retrievals are listed in Table \ref{table:planets}.

Our results are presented as follows. In Section \ref{results:water abundance}, we first discuss the most constrained parameter given the data: the H$_2$O abundance. We then briefly consider other chemical species and their abundances in Section \ref{results:other species}. We explore potential trends among the planetary parameters, H$_2$O abundances, and cloud/haze properties in Section \ref{results:clouds}. We emphasize that the availability of optical HST STIS data for all planets allows a robust determination of H$_2$O abundances, significantly reducing the degeneracies that arise from consideration of HST WFC3 data alone. This is exhibited in Section \ref{optical_data}.

\begin{figure*}
\centering
\includegraphics[scale=0.57]{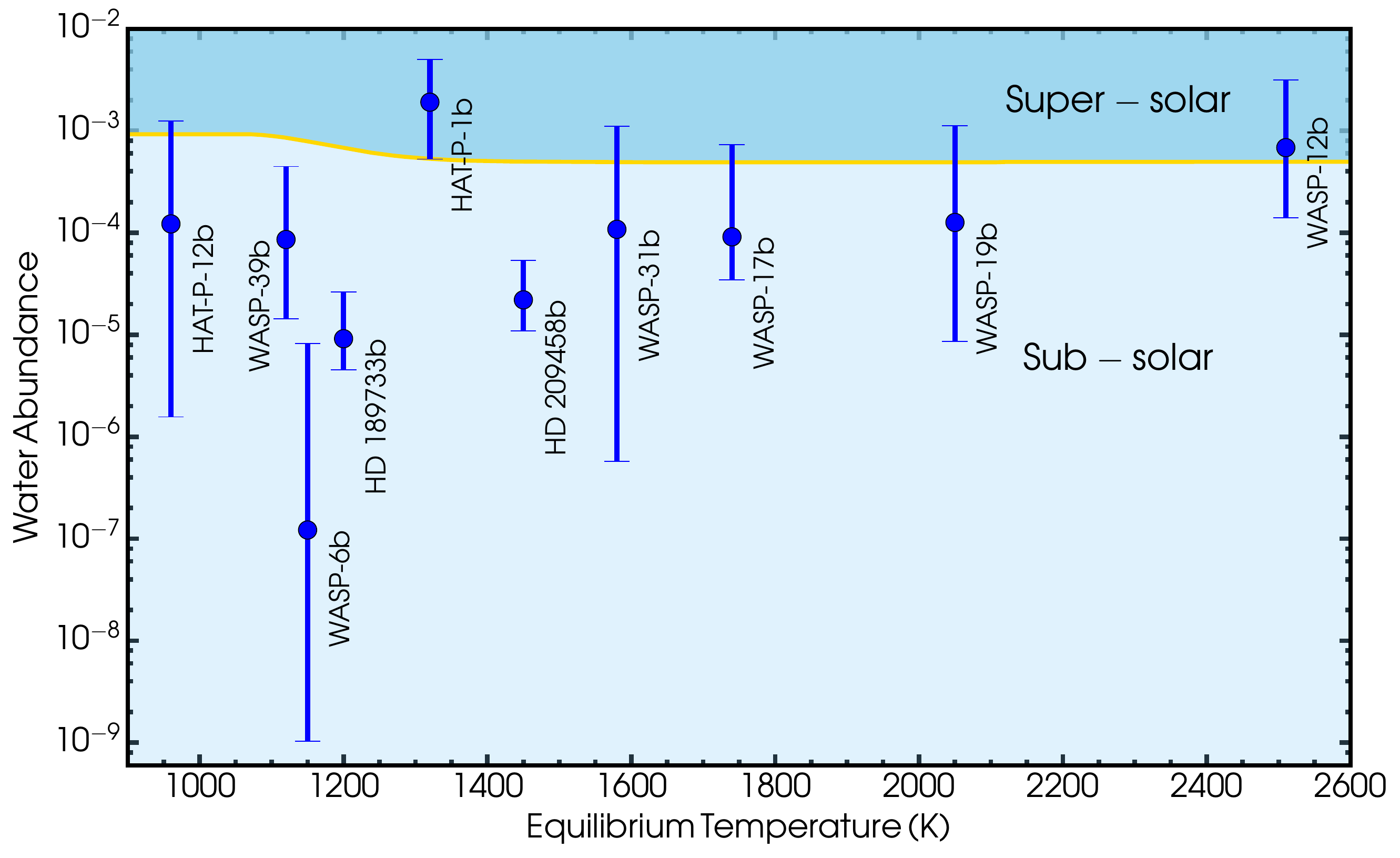}%
\caption{The retrieved H$_2$O volume mixing ratios and their associated $1\sigma$ ranges for the hot Jupiter sample. The planets are consistent with sub-solar H$_2$O abundances within 1$\sigma$, excepting HAT-P-1b. The gold line shows the volume mixing ratio of water vapour calculated from solar elemental abundances \citep{madhu12} and ranges from $\sim$$10^{-3}$ at $T$ $\lesssim$ 1,200 K to $5 \times 10^{-4}$ at higher temperatures.}
\label{fig:h2o_abunds}
\end{figure*}

\begin{table*}
\centering 
\begin{tabular}{c c c c}
\hline
\hline
\rule{0pt}{3ex}  Planet & Log($X_{\mr{H_2O}}$$_{-1\sigma}^{+1\sigma}$) & Normalised Abundance ($X^{\odot}_{\mr{H_2O}}$)$_{-1\sigma}^{+1\sigma}$ & Detection Significance\\ [1ex]
\hline
\rule{0pt}{3ex}
 HAT-P-12b & $-3.91^{+1.01}_{-1.89}$ & $0.133^{+1.226}_{-0.131}$ & N/A\\[1ex]
 WASP-39b & $-4.07^{+0.72}_{-0.78}$ & $0.10^{+0.42}_{-0.08}$ & 7.00$\sigma$\\[1ex] 
 WASP-6b & $-6.91^{+1.83}_{-2.07}$ & $0.000156^{+0.014}_{-0.000155}$ & N/A\\[1ex]
 HD 189733b & $-5.04^{+0.46}_{-0.30}$ & $0.018^{+0.035}_{-0.009}$ & 5.60$\sigma$\\[1ex]
 HAT-P-1b & $-2.72^{+0.42}_{-0.56}$ & $3.58^{+5.84}_{-2.60}$ & 3.50$\sigma$\\[1ex]
 HD 209458b & $-4.66^{+0.39}_{-0.30}$ & $0.04^{+0.06}_{-0.02}$ & 7.06$\sigma$\\[1ex]
 WASP-31b & $-3.97^{+1.01}_{-2.27}$ & $0.218^{+2.01}_{-0.217}$ & 2.05$\sigma$\\[1ex]
 WASP-17b & $-4.04^{+0.91}_{-0.42}$  & $0.19^{+1.32}_{-0.12}$ & 3.22$\sigma$\\[1ex]
 WASP-19b & $-3.90^{+0.95}_{-1.16}$  & $0.26^{+2.03}_{-0.24}$ & 2.89$\sigma$\\[1ex]
 WASP-12b & $-3.16^{+0.66}_{-0.69}$ & $1.40^{+4.99}_{-1.11}$ & 5.73$\sigma$\\[1ex]
\hline
\hline
\end{tabular}
\parbox{10.4cm}{\caption{Terminator H$_2$O abundances, solar-normalised H$_2$O abundances, and detection significances. The normalised H$_2$O abundances are relative to the solar values shown by the gold line in Figure \ref{fig:h2o_abunds}. Detection significances are given for Bayes factors greater than 2.64 (i.e. 2$\sigma$).}
\label{tab:h2o}}
\end{table*}

\subsection{Chemistry}

\subsubsection{H$_2$O Abundance}\label{results:water abundance}

The H$_2$O abundances are shown against planetary equilibrium temperature in Figure \ref{fig:h2o_abunds}. The reference `solar' H$_2$O indicates the H$_2$O abundance as a function of temperature expected in hot Jupiter atmospheres with solar elemental abundances at a pressure of 1 bar \citep{asplund09, madhu12}. This is shown by the gold line in Figure \ref{fig:h2o_abunds}. For the majority of hot Jupiters considered in this work (i.e. those with equilibrium temperatures above $\sim$1,300 K), H$_2$O is expected to contain $\sim$50\% of the total available oxygen \citep{madhu12} such that the solar water abundance is a constant and is $X^{\odot}_{\mr{H_2O}} = \frac{1}{2}\frac{\mr{O}}{\mr{H_2}}\big{|_{\odot}} X^{\odot}_{\mr{H_2}}$, where $\frac{\mr{O}}{\mr{H_2}}\big{|_{\odot}} = 2 \frac{\mr{O}}{\mr{H}}\big{|_{\odot}}$ and log($X^{\odot}_{\mr{H_2O}}$) = $-3.3$.

The planets with the most precise observations in the HST WFC3 G141 bandpass -- HD 189733b, HD 209458b, and WASP-12b -- have retrieved $\mr{log(X_{\mr{H_2O}})}$ abundances of $-5.04^{+0.46}_{-0.30}$, $-4.66^{+0.39}_{-0.30}$, and $-3.16^{+0.66}_{-0.69}$, respectively. The two hot Jupiters with the highest quality observations (HD 189733b and HD 209458b) show the most statistically significant H$_2$O depletion and are consistent with those of previous studies \citep{madhu14b,barstow17,macdonald17_209}. The inferred H$_2$O abundance of WASP-12b is consistent to 1$\sigma$ with that of another study \citep{kreidberg15}. HAT-P-1b, WASP-31b, WASP-17b, WASP-19b, and WASP-12b contain relative abundances of $3.58^{+5.84}_{-2.60}\times$, $0.218^{+2.01}_{-0.217}\times$, $0.19^{+1.32}_{-0.12}\times$, $0.26^{+2.03}_{-0.24}\times$, $1.40^{+4.97}_{-1.11}\times$ solar which are consistent with sub-solar and super-solar concentrations to within $1\sigma$. The retrieved water abundances for the planets with the lowest equilibrium temperatures, HAT-P-12b and WASP-39b, are $-3.91^{+1.01}_{-1.89}$ and $-4.07^{+0.72}_{-0.78}$, corresponding to $0.133^{+1.226}_{-0.131}\times$ and $0.10^{+0.42}_{-0.08}\times$ the solar abundance. The estimated water abundance for WASP-39b is inconsistent with that of another study \citep{wakeford18} which used different WFC3 data than in the present work. WASP-6b has the lowest derived abundance, consistent with a non-detection, owing largely to the lack of a HST WFC3 spectrum.

The H$_2$O abundances, the solar-relative abundances, and the detection significances of water vapour are listed in Table \ref{tab:h2o}. Considering the H$_2$O abundance estimates for all the planets as an ensemble the representative mixing ratio is log(X$_{\mr{H_2O}}$) = $-4.20^{+0.20}_{-0.17}$ or about $0.07^{+0.04}_{-0.02}\times$ solar. This ensemble value is inconsistent with a solar or super-solar abundance at $5.37\sigma$ assuming the objects come from the same population. {Importantly, we have tested for effects of stellar heterogeneity using the CPAT model \citep{rackham17, pinhas18_Aura} and find no significant changes in the water abundances.} We find no clear quantitative correlation between the H$_2$O abundance and $T_{\mr{eq}}$.

\subsubsection{Other chemistry}\label{results:other species}

We also constrain the presence and abundance of alkali absorbers in addition to H$_2$O. Table \ref{tab:add_species} shows these elements which show clear modes in the retrieved posterior distributions with significances above $2\sigma$. HD 189733b shows a confident detection of Na corresponding to a significance of 5.01$\sigma$ (Bayes factor of $4.7 \times 10^{4}$) and a sub-solar value of $-7.77^{+1.64}_{-0.87}$. We detect potassium in WASP-6b at $2.67\sigma$ confidence with a constrained abundance of $-5.53^{+2.01}_{-1.85}$, consistent with the solar value. While there is a clear potassium signature in the spectrum of WASP-31b, the fidelity of this data point has recently been called into question \citep{gibson17} and therefore WASP-31b is not included in Table \ref{tab:add_species}. The retrieved posteriors for WASP-39b show evidence for CH$_4$ and CO$_2$ and yet may be due to a possible systematic offset between the HST and Spitzer data. Evidence of nitrogen-bearing molecules in these planetary atmospheres is examined in \citet{macdonald17_Nchem}.

\begin{table}
\centering 
\caption{Retrieved atomic species with detection significances above 2$\sigma$.}
\label{tab:add_species}
\begin{tabular}{c c c c}
\hline
\hline
\rule{0pt}{3ex}  Planet & Species & Detection Significance & Abundance\\ [1ex]
\hline
\rule{0pt}{3ex}
 
 \\
 
 WASP-39b & \parbox{1cm}{\centering Na \\ \centering K} & \parbox{1cm} {\centering 3.41$\sigma$ \\ \centering 3.62$\sigma$} & \parbox{1.2cm} {$-3.86^{+1.31}_{-1.36}$ \\ $-4.22^{+1.25}_{-1.12}$} \\[1ex] 
 
 \\
 
 WASP-6b & K & 2.67$\sigma$ & ${-5.53}^{+2.01}_{-1.85}$\\[1ex]

 \\
 
 HD 189733b & Na & 5.01$\sigma$ & ${-7.77}^{+1.64}_{-0.87}$ \\[1ex]
 
 \\
 
 HAT-P-1b & Na & 2.22$\sigma$ & $-8.44^{+1.45}_{-2.12}$\\[1ex]
 
\hline
\hline
\end{tabular}
\end{table}

\begin{figure*}
\centering
\includegraphics[scale=0.65]{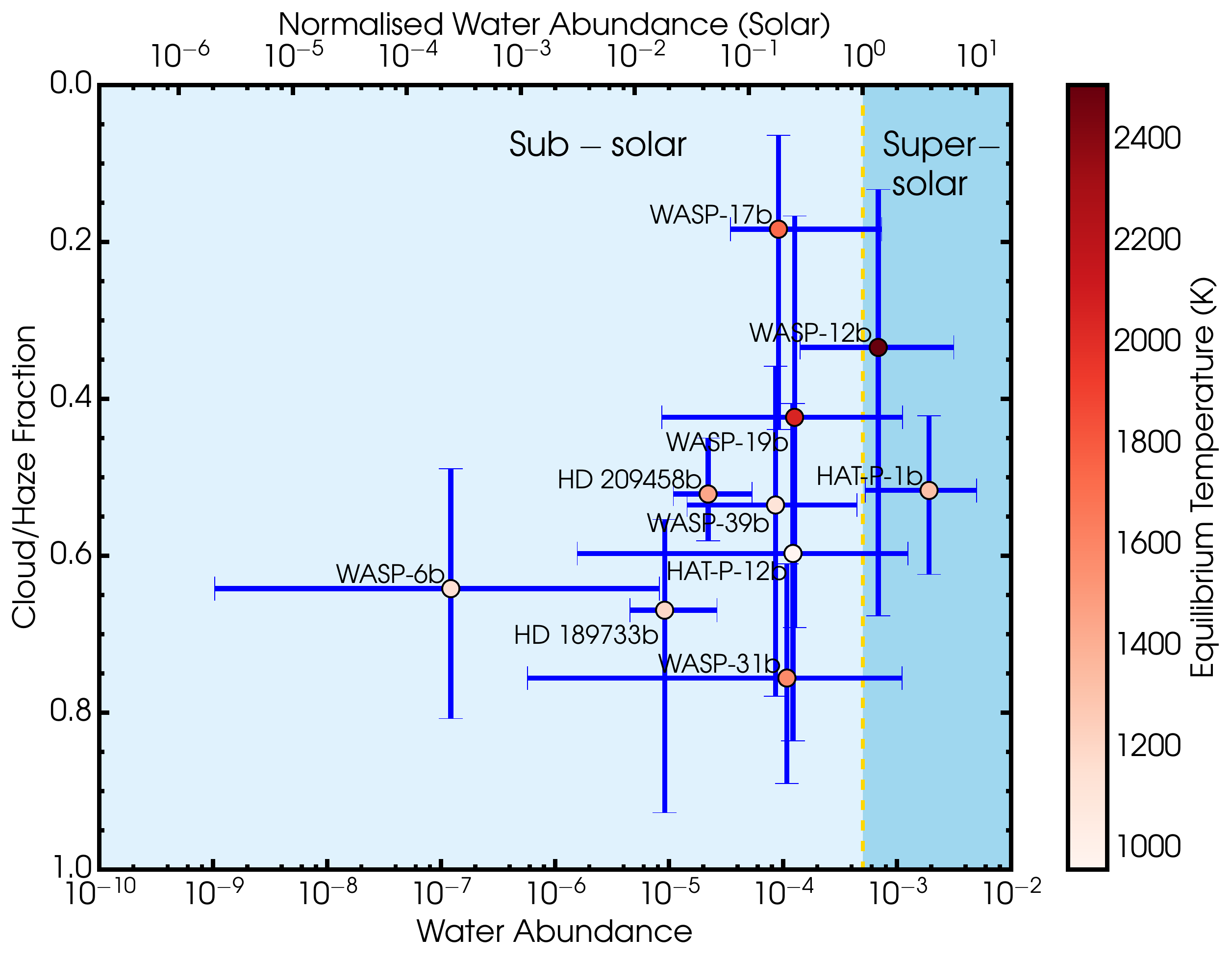}%
\caption{Terminator cloud/haze fractions, H$_2$O abundances, and equilibrium temperatures of the hot Jupiter sample. The dashed gold line represents the solar water abundance at high temperatures (i.e. $5\times 10^{-4}$).}
\label{fig:sing_figure1}
\end{figure*}

\begin{figure*}
\centering
\includegraphics[scale=0.5]{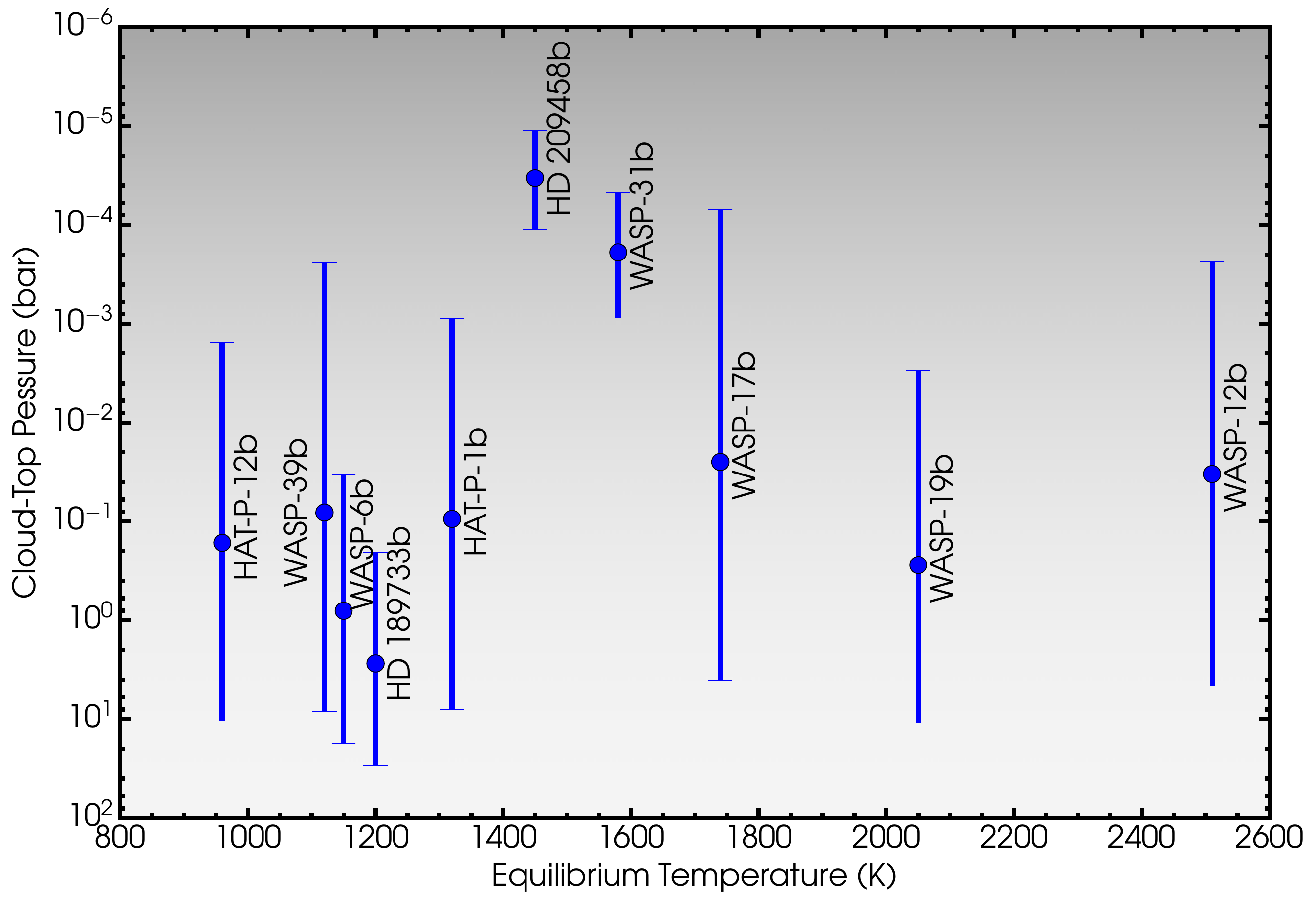}%
\caption{Top pressure of grey clouds versus planetary equilibrium temperature for the hot Jupiter ensemble.}
\label{fig:comparison1}
\end{figure*}

\begin{figure*}
\centering
\includegraphics[scale=0.5]{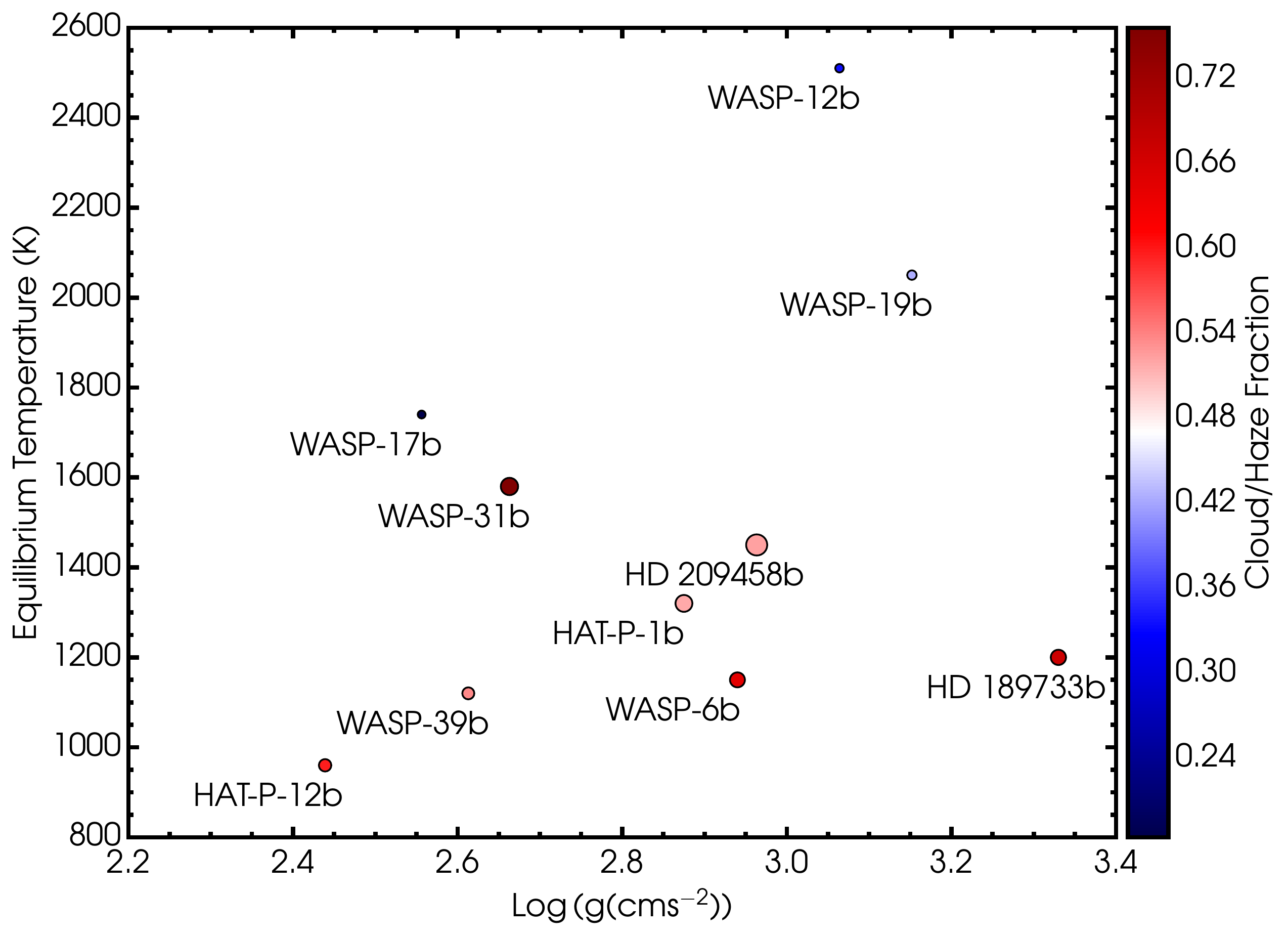}%
\caption{Cloud fractions $\bar{\phi}$ as a function of $T_{\mr{eq}}$ and planetary gravity. The marker size is inversely proportional with the retrieved $1\sigma$ confidence on the cloud fraction, such that finer constraints are represented with larger circles.}
\label{fig:comparison2}
\end{figure*}

\subsection{Exploring Trends with Planetary Parameters, H$_2$O Abundances, and Cloud/Haze Properties}\label{results:clouds}

We have carried out an extensive exploration of potential trends among planetary parameters, H$_2$O abundances, and cloud/haze properties. We have investigated 35 combinations of pairs and triplets over the parameters $T_{\mr{eq}}$, $M_p$, $g$, $X_{\mr{H_2O}}$, $\bar{\phi}$, and $P_{\mr{cloud}}$, resulting in no clear correlations. However, we present three parameter spaces and compare with analogous presentations in the literature \citep{sing16,barstow17,stevenson16}. An extensive comparison with the methodologies and results of \citet{sing16} and \citet{barstow17} is presented in Section \ref{conclusions}.

The terminator cloud/haze fractions ($\bar{\phi}$) versus H$_2$O abundances with $T_{\mr{eq}}$ as a third dimension are shown in Figure \ref{fig:sing_figure1}. We do not find a clear trend between $\bar{\phi}$ and H$_2$O, and the cloud fractions for the planetary sample do not follow the clear-to-cloudy/hazy trend suggested by the spectral survey \citep{sing16}. The median cloud/haze fractions for WASP-12b and WASP-6b imply fewer aerosols and those for HD 209458b, WASP-31b, and HAT-P-12b imply more aerosols compared to a previously suggested order \citep{sing16}. Moreover, the derived grey cloud-top pressures (P$_{\mr {cloud}}$) versus $T_{\mr{eq}}$ are shown in Figure \ref{fig:comparison1}. HD 189733b and WASP-6b likely have clouds composed of large particles deep in the atmospheres below $\sim$$10^{-1}$ bar. On the other hand, opaque clouds with maximal cloud-top pressures of 0.1 mbar are found for HAT-P-12b, WASP-39b, HAT-P-1b, WASP-17b, WASP-19b, and WASP-12b but are relatively unconstrained. HD 209458b and WASP-31b have precise constraints of $\sim$0.5 dex on the cloud-top pressures and the cloud-tops lie above 1 mbar. Overall, we find no correlation between T$_{\mr{eq}}$ and P$_{\mr {cloud}}$. However, planets with low cloud-top pressures (below $\sim$1 mbar) span equilibrium temperatures of 1400-1600 K.

Finally, the space of $T_{\mr{eq}}$, $g$, and $\bar{\phi}$ is shown in Figure \ref{fig:comparison2}. The lack of a clear correlation among these parameters is unlike previously suggested \citep{stevenson16}. For the equilibrium temperatures spanned in our work ($T_{\mr{eq}} > 700$ K), \citet{stevenson16} suggested that planets with $\mr{log}$$(g[\mr{cms^{-2}}])$ greater than 2.8 are cloud-free whilst those below 2.8 should host a significant cloud fraction. In contrast, we find no such division, similar to a conclusion from a recent study \citep{barstow17} of the spectral survey \citep{sing16}. This difference exists for at least two reasons. Firstly, some of the G141 observations used in \citet{stevenson16} are different than those we retrieve. Secondly, \citet{stevenson16} explains the H$_2$O feature amplitude with reference to clouds alone, assuming no variations over the H$_2$O abundance. This assumption is too restrictive since a low-amplitude feature can imply peculiarly low water abundance with minimal cloud coverage and/or high $P_{\mr{cloud}}$ or a high water abundance with a significant cloud fraction and/or low $P_{\mr{cloud}}$. On the other hand, the relatively clearer atmospheres of WASP-12b, WASP-19b, and WASP-17b shown in Figure \ref{fig:comparison2} support suggestions of hotter ($T_{\mr{eq}} \gtrsim 1700\,\mr{K}$) close-in planets harbouring less cloudy atmospheres \citep{heng16, tsiaras17, liang04}.

\subsubsection{Metallicity and Formation Conditions}

\begin{figure*}
\centering
\includegraphics[scale=0.65]{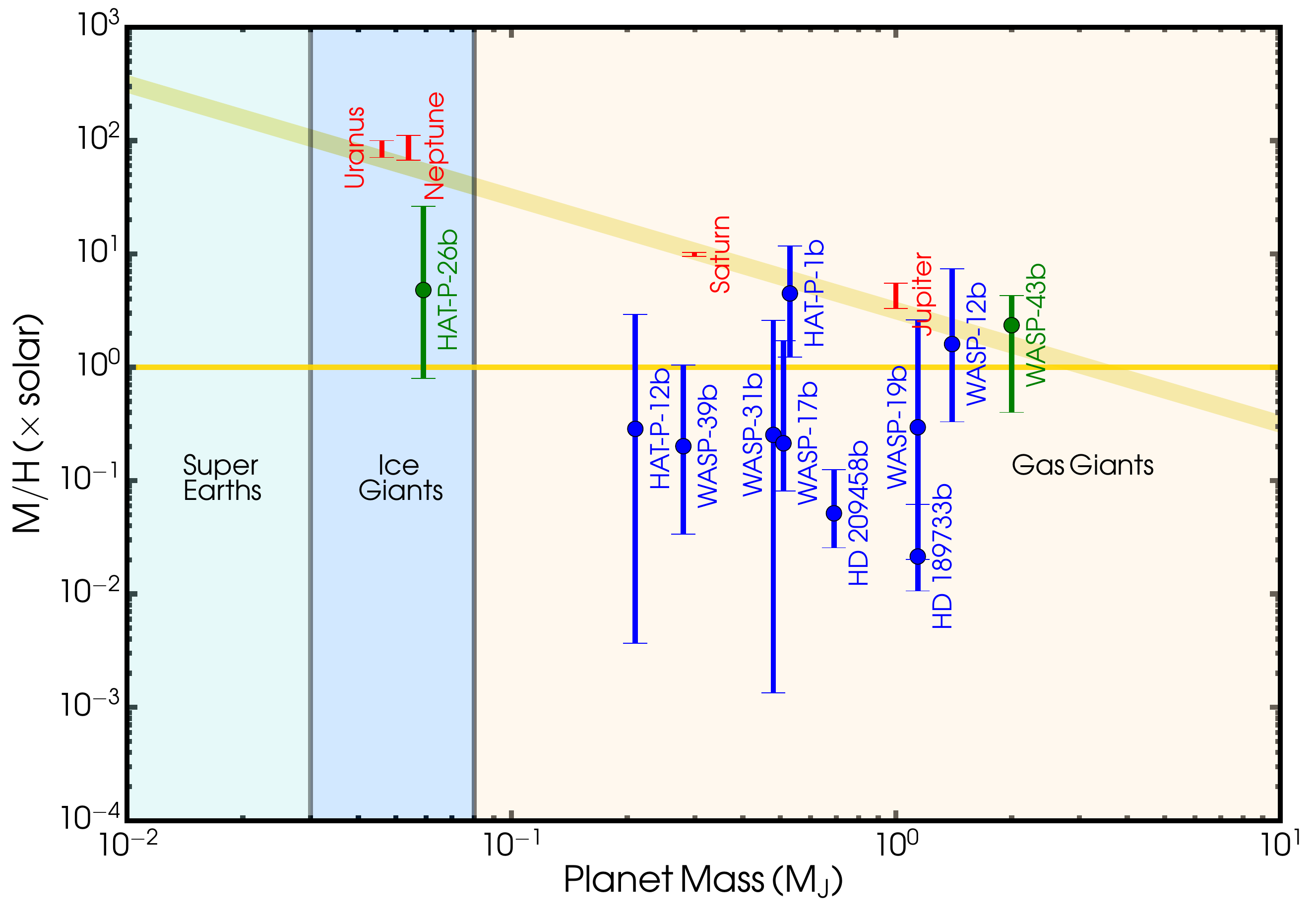}%
\caption{Atmospheric metal abundances as a function of planetary mass for the hot Jupiter ensemble. The planets in the current study are shown in blue, along with WASP-43b \citep{kreidberg14} and HAT-P-26b \citep{wakeford17} from previous studies shown in green. The solar system giant planets are shown in red. The solar metallicity reference is shown by the horizontal gold line. WASP-6b is excluded due to lack of WFC3 observations. The exoplanetary atmosphere metallicities M/H of the giant exoplanets in our sample are derived from the H$_2$O volume mixing ratio, such that M represents O, and assuming that half of the oxygen is in H$_2$O as expected for a solar C/O ratio at high temperatures \citep{madhu12}. The metallicities of solar system planets are derived from the CH$_4$ abundances, such that M represents C.} 
\label{fig:met_mass}
\end{figure*}

The retrieved set of H$_2$O abundances provide initial clues on metal abundances in the exoplanetary atmospheres. Figure \ref{fig:met_mass} shows the space of atmospheric metallicity versus planet mass for the hot Jupiter sample in addition to previously reported metallicities of WASP-43b \citep{kreidberg14}, HAT-P-26b \citep{wakeford17}, and the solar system gas giants. The atmospheric metallicities are estimated from molecular species with well-determined abundances since knowledge of the full inventory of metals is limited by observational capabilities. The atmospheric metallicities for the hot Jupiter sample are calculated assuming that half of the oxygen is in the estimated H$_2$O and the remaining half is in CO, in accordance with expectations for a solar C/O ratio in chemical equilibrium at high temperatures \citep{madhu12}. The atmospheric metallicities of the four solar system gas giants are determined from the abundance of methane which contains most of the carbon at their low temperatures \citep{atreya16}. The abundance of oxygen in the atmospheres of solar system planets cannot be incorporated into a metallicity estimate since much of it is condensed in deep-lying water clouds.

We emphasize that there are limitations to the illustration in Figure \ref{fig:met_mass} due to the different molecules used to represent the metallicities as well as using one molecular species as a metallicity descriptor. In principle, a high C/O ratio (e.g. C/O$\sim$1) can lead to significantly sub-solar H$_2$O. Therefore, the low H$_2$O abundances would indicate a low metallicity and/or a high C/O ratio \citep{madhu14b}. The metallicity proxy for solar system planets shows a decreasing trend with increasing planetary mass. The sub-solar hot Jupiter metallicities suggest a weak trend that is different from that of the solar system gas giants.

The inferred oxygen abundances provide initial clues into the formation and migration scenarios of these hot Jupiters when considering the metallicities of their host stars. The host stars of the majority of these hot Jupiters have O/H abundances in excess of the solar value \citep{teske14,brewer17}. This implies that the O/H ratios in the majority of the planetary atmospheres are sub-stellar as well as being sub-solar. Hot Jupiter atmospheres are expected to possess super-stellar oxygen abundances if they are formed through core-accretion followed by migration within the disk \citep{madhu14a,mordasini16}, as also suggested for Jupiter based on its super-solar abundances in several elements \citep{owen99, atreya05, mousis12}.

On the other hand, the generally sub-solar and sub-stellar oxygen abundances found in these atmospheric spectra suggest a general scenario in which the hot Jupiters form far from their stars with efficient gas accretion and relatively inefficient solid planetesimal accretion \citep{madhu14b,madhu16a} since the gaseous O/H abundance is low in outer regions of protoplanetary disks due to successive condensation fronts of O-rich species \citep{oberg11}. Subsequent impulse inwards through gravitational interactions with other bodies in the system by disk-free migration may have brought many of these hot Jupiters to their present locations \citep{rasio96}. Such disk-free migration could lead to low oxygen abundances irrespective of formation mechanisms, either core-accretion or gravitational instability \citep{madhu14a}. Alternatively, some of the hot Jupiters may have formed through pebble accretion and migrated inward with or without the disk but without significant erosion of the core \citep{madhu17,booth17}. The suggested importance of disk-free migration implied from the general trend of low oxygen abundances is consistent with proposals in other studies of the prevalence of disk-free migration based on dynamical properties of hot Jupiters and their environments \citep[e.g.][]{nelson17,burcalassi16}.

\subsection{Importance of Optical Data}\label{optical_data}

\begin{figure*}
    \includegraphics[width=\textwidth]{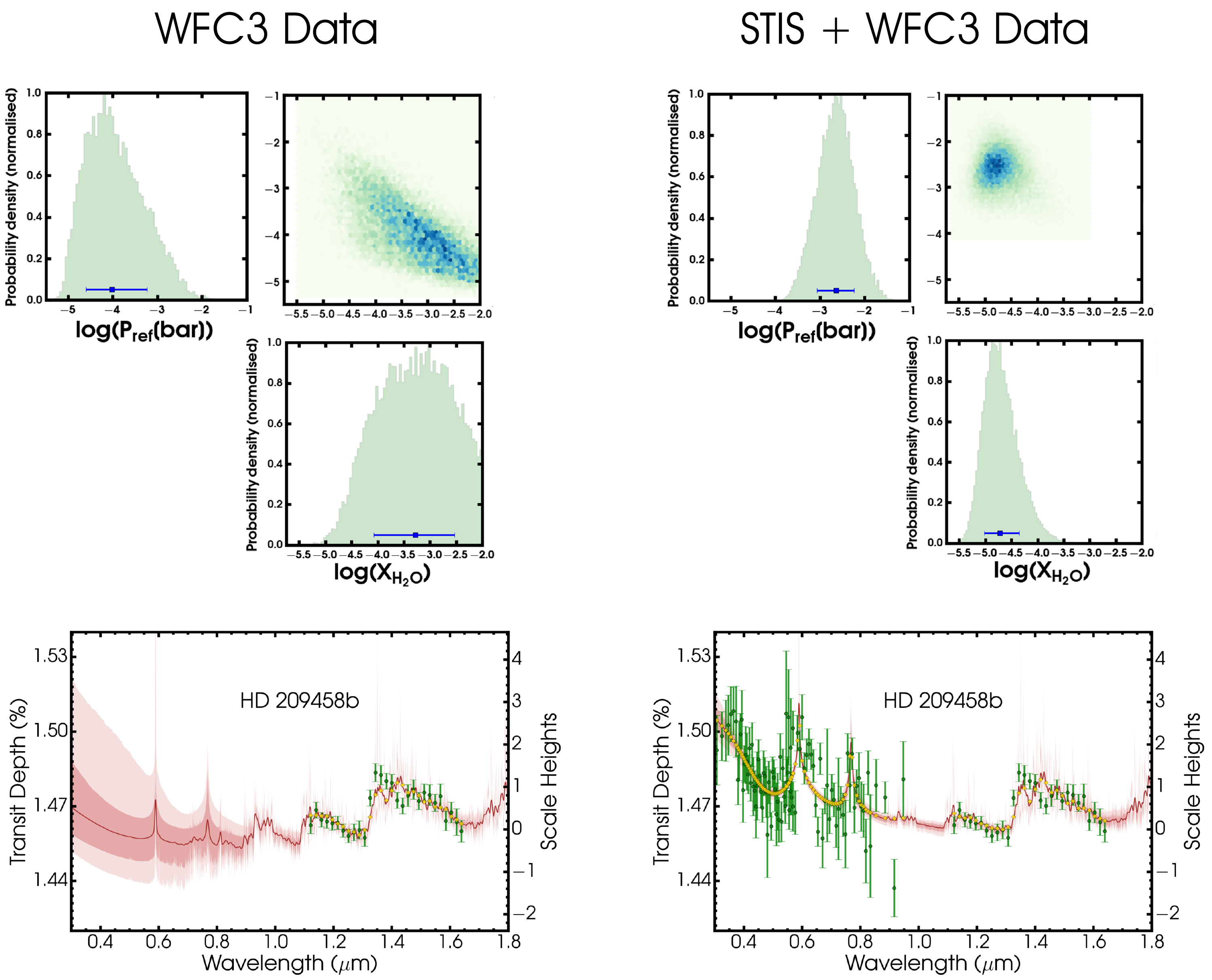}
    %\caption{Picture 2}
    %\label{fig:2}
\caption{Demonstration of the use of optical data to precisely determine $P_{\mr{ref}}$ and $X_{\mr{H_2O}}$. HD 209458b is used for illustration. One retrieval (left column) inverts the HST WFC3 data and another retrieval (right column) includes HST STIS data in the optical in addition to WFC3 data. The juxtaposition demonstrates that data in the optical range break the degeneracy between $P_{\mr{ref}}$ and $X_{\mr{H_2O}}$.}
\label{fig:break_norm_degen} 
\end{figure*}

The inclusion of optical data is essential in the interpretation of the hot Jupiter spectra, especially in the estimation of H$_2$O abundances. The interpretation of HST WFC3 data alone introduces a well known ambiguity between the reference pressure $P_{\mr{ref}}$ and the water abundance $X_{\mr{H_2O}}$ \citep[e.g., see][]{griffith14}. The use of HST STIS data to infer a reliable constraint on the reference pressure and the water abundance is illustrated in Figure \ref{fig:break_norm_degen}. In the retrieval of HST WFC3 data alone the line of ambiguity between $P_{\mr{ref}}$ and $X_{\mr{H_2O}}$ spreads over two orders of magnitude and the inferred $X_{\mr{H_2O}}$ are mostly contained above $10^{-4}$. The same degeneracy disappears when optical HST STIS data are included in the retrieval; the joint posterior between $P_{\mr{ref}}$ and $X_{\mr{H_2O}}$ shows a well-localised set of solutions with no correlative trend. Moreover, the $X_{\mr{H_2O}}$ values are constrained three times better and are contained below $10^{-4}$. The juxtaposition in Figure \ref{fig:break_norm_degen} illustrates, in no unclear terms, that the broad ambiguity between $P_{\mr{ref}}$ and $X_{\mr{H_2O}}$ that exists for near-infrared WFC3 data alone collapses through the use of optical data, in this case HST STIS observations.

\section{Discussion and Conclusions}\label{conclusions}

Here we compare our work with two previous studies: the spectral survey of \citet{sing16} that first reported the observations and \citet{barstow17} that conducted initial retrievals on those datasets. The differences between our work and these previous studies lie in both the modeling and retrieval approaches, as well as the results. At the outset, the key aspect of our present work is the ability to derive statistical constraints (i.e. estimates with non-Gaussian error bars), along with full posterior distributions, for all the atmospheric parameters concerned. This allows one to pursue comparative exoplanetology for a sizeable exoplanet sample.

The differences in H$_2$O abundances and cloud/haze properties between our findings and those of \citet{sing16} can be attributed to several crucial factors. Firstly, the conclusions of the latter study were reached through a comparison of the data with equilibrium models of hot Jupiter atmospheres computed over a pre-determined grid in  metallicity and temperature, as opposed to a full retrieval as in the present study. Given the limited number of models in the grid and the equilibrium assumption, it was not possible to explore the model space adequately nor could the atmospheric properties be estimated statistically. This meant that it was necessary to invoke empirical metrics to qualitatively assess the contributions of the H$_2$O abundance and clouds/hazes in the atmospheres. For example, the amplitude of the H$_2$O feature at $\sim$1.4 $\mu$m was suggested to correlate with the difference between the planetary radius in the optical and infrared \citep{sing16}. However, this correlation was based on cloudy/hazy models which assumed solar and super-solar H$_2$O abundances, without an exploration of cloudy/hazy atmospheres with sub-solar water abundances as well as sub-solar cloudy/hazy opacities (see their Figure 3). As such, their conclusions were unable to reflect the full range of possible solutions, as confirmed in a follow-up study by \citet{barstow17} (discussed below) which found predominantly sub-solar H$_2$O solutions. Our study instead uses a comprehensive retrieval of all the datasets using a Bayesian nested sampling approach, enabling a detailed exploration of the entire multi-dimensional space of atmospheric parameters. We thereby assume no a priori values for the H$_2$O abundances and cloud/haze properties.

There are also several differences between our study and that of \citet{barstow17}. The differences are of three kinds: model parameterizations, statistical sampling algorithms, and reported estimates. We here discuss each of these in turn.

There are several important differences in the model parameterizations. The main aspect of the \citet{barstow17} study is that it is effectively a grid-based retrieval, in the sense that a separate retrieval is conducted for each assumption of a cloud prescription. Importantly, the study treats two kinds of cloud models separately: a model with Rayleigh-like  slopes (i.e. our hazes) and one with grey opacity (i.e. our clouds). On the other hand, our cloud/haze model simultaneously accounts for small and large particle sizes (see equation (4)). An important distinction also lies in the aerosol parameterisation. Our study includes a grey cloud overlain by a haze such that both types of aerosols are retrieved for a planet, although the retrieved parameters could indicate no hazes (i.e. $a=1$ and $\gamma = -4$) and/or no grey clouds (i.e. $P_{\mr{cloud}}$ $\gtrsim$ 10 bar). On the other hand, \citet{barstow17} allows for the presence of a vertically-finite cloud or haze deck at variable locations in the atmosphere. In some cases (e.g. HD 189733b) where our study retrieves a deep cloud layer with an extensive overlying haze layer, \citet{barstow17} retrieve a decade-confined haze layer confined to low pressures. These two different interpretations, however, produce similar fits to the spectrum.

An additional crucial difference in the aerosol models is that the present study accounts for inhomogeneous clouds and hazes along the planetary limb. The inclusion of patchy clouds is important since the shape of the H$_2$O feature in the WFC3 bandpass is sensitive to the degree of cloud contribution \citep[see e.g.][]{line16}. These differences in modeling are likely another principal element responsible for discrepancies in the cloud and haze properties between the present work and that of \citet{barstow17}. Second, \citet{barstow17} assumes an isotherm for the observable region of the atmosphere (pressures below $0.1$ bar) whereas we allow a fully general temperature profile that allows for any temperature gradient. Third, there is a significant difference in the treatment of opacity in the radiative transfer. Whereas the previous study assumes the correlated-$k$ distribution for opacities, our study uses opacity sampling from very high resolution cross-sections for the different molecules. This difference in opacity treatments could potentially influence the retrievals. However, generally similar water abundance estimates between the two methods is encouraging. Fifth, the cloud properties and temperature profiles in \citet{barstow17} were pre-defined on a grid, whereas the atmospheric parameters in our study are sampled continuously over the entire parameter space.

There are also important differences in the statistical inference methods. \citet{barstow17} uses an optimal estimation (OE) sampler whereas our analysis uses a Bayesian nested sampling algorithm, {\sc MultiNest}. First, while OE has been shown to be accurate for very high resolution data, its accuracy has been shown to be limited for low resolution spectrophotometry \citep{line13} as is relevant in the present case. On the other hand, the {\sc MultiNest} Bayesian analysis approach has been shown to be more accurate for the quality of hot Jupiter data considered here since it is able to explore parameter spaces with multi-modal solutions \citep{feroz08,feroz13}. Second, the OE sampler in \citet{barstow17} requires assuming Gaussian-distributed uncertainties in the model parameters whereas no such assumption is required in the present Bayesian approach. This is significant since the derived parameter posteriors in our study indeed show mostly non-Gaussian distributions. Third, given the efficiency of our modeling and retrieval approach a much larger volume of the high-dimensional model parameter space is explored for each planet (i.e., $\sim$$5 \times 10^{6}$ models) while the approach in \citet{barstow17} was limited to 3,600 model evaluations per planet.

Finally, there are important differences regarding the nature of reported parameter estimates. First, \citet{barstow17} does not provide statistical estimates on the abundances and other atmospheric properties but instead provides parameter values of a select set of model fits to the data. On the other hand, our analysis provides statistical limits, along with full posteriors, for all the atmospheric parameters through consideration of all model evaluations. Hence, our estimated median values and 1$\sigma$ uncertainties are obtained through marginalization over the full posteriors. Second, the cloud/haze properties and temperatures profiles in \citet{barstow17} were pre-defined on a grid and hence joint constraints on these properties and chemistry are not possible. Our approach allows for complete marginalization over all parameters and hence provides joint statistical correlations between all atmospheric parameters, thereby enabling a more extensive illumination of the atmospheres.

The considerations above naturally lead to differences between our results and those of the \citet{barstow17} study, particularly on the nature of clouds. \citet{barstow17} finds three planets -- WASP-39b, HD 209458b, and WASP-31b -- are best fit by grey opacity clouds with top pressures of 10$^{-5}$ bar, 0.01 bar, and 0.1 bar, respectively. Figure \ref{fig:comparison1} shows the retrieved cloud-top pressures for HD 209458b and WASP-31b are smaller by $\sim$2 dex and that of WASP-39b is larger by $\sim$4 dex. A comparison with other planets in the sample is not possible since the cloud-top pressures in the previous study are quoted for pure Rayleigh clouds \citep{barstow17}, i.e. only for particles of small sizes.

In addition, a trend is suggested in the analysis of \citet{barstow17} such that planets with equilibrium temperatures spanning 1300 K to 1700 K possess grey clouds that are confined deep in their atmospheres below $\sim$10 mbar \citep{barstow17}. On the other hand, our analysis finds that HAT-P-1b ($T_{\mr{eq}} = 1320$ K), HD 209458b ($T_{\mr{eq}} = 1450$ K), and WASP-31b ($T_{\mr{eq}} = 1580$ K) are consistent with high-altitude grey cloud-top pressures (see Figure \ref{fig:comparison1}). WASP-17b and WASP-12b, with equilibrium temperatures above 1700 K, have also been suggested to have good evidence for small-particle aerosols at high atmospheric altitudes \citep{barstow17}. We find that a model without hazes or clouds composed of small particles provides statistically comparable fits to the spectra of WASP-17b and WASP-12b, and is suggestive of weak evidence for small-particle aerosols in their atmospheres.

The discussion above illustrates there are differences in the retrieved cloud properties between our analysis and that of \citet{barstow17} which may be attributed to the different approaches of cloud/haze modelling. In spite of these differences, we emphasize that the estimated water abundances in the hot Jupiter atmospheres are generally in good agreement and show a trend towards sub-solar values. Planets for which similar WFC3 data are used (HD 189733b, HD 209458b, WASP-31b, WASP-17b, and WASP-19b) illustrate that the H$_2$O abundances can be robust to different aerosol treatments in retrieval studies.

Beyond the key advancements in the modelling approach, the Bayesian inference method, and the results, the major contribution of our work is to provide detailed statistical estimates of important atmospheric properties for a sizable sample of hot Jupiters. All in all, these estimates will prove invaluable for comparative planetology, across both  exoplanetary and solar system giant planets and for understanding their formation pathways.

In summary, we have carried out a comprehensive atmospheric study of transmission observations of a sizeable giant exoplanet sample contained in \citet{sing16}. Through a homogeneous Bayesian retrieval analysis of the planetary spectra, we determine statistically robust estimates of various atmospheric parameters including the H$_2$O abundances, cloud/haze properties, other chemical abundances, and temperature profiles. In particular, we find that all the planetary atmospheres are consistent in harbouring sub-solar H$_2$O abundances within 1$\sigma$ with the exception of HAT-P-1b. The planets with the most precise observations in the HST WFC3 G141 bandpass -- HD 189733b, HD 209458b, and WASP-12b -- are constrained to have H$_2$O abundances of $0.018^{+0.035}_{-0.009}\times$, $0.04^{+0.06}_{-0.02}\times$, and $1.40^{+4.97}_{-1.11}\times$ solar, respectively. We find a continuum over cloud and haze contributions as suggested in recent studies \citep{barstow17, sing16}, although the details are different than suggested therein.

The lack of a clear correlation among various properties of the atmospheres and macroscopic properties of the planets is consistent with a unique and detailed evolutionary history for each giant exoplanet. Nevertheless, in light of the host stars' solar or super-solar O/H metallicities and the generally sub-solar O/H abundances, the majority of close-in hot giant exoplanets are suggested to form beyond the H$_2$O ice-line with subsequent dynamical interaction and disk-free migration to their present environments.

\section*{Acknowledgements}

AP is grateful for research support from the Gates Cambridge Trust. NM, SG, and RM acknowledge support from the Science and Technology Facilities Council (STFC), UK. This research has made use of the SVO Filter Profile Service supported from the Spanish MINECO through grant AyA2014-55216. We thank the contributors of the Python Software Foundation, the Open Science Framework, and NASA's Astrophysics Data System.

\bibliographystyle{mnras}
\bibliography{references} 
\appendix
\section{COMPLETE RETRIEVED PARAMETERS}

\begin{sidewaystable*}
\caption{Retrieved chemical abundances of the hot Jupiter sample.}
\label{tab:retvals1}
\centering
\begin{threeparttable}
\begin{tabular}{K{1.2cm} K{1.65cm} K{1.65cm} K{1.65cm} K{1.65cm} K{1.65cm} K{1.65cm} K{1.65cm} K{1.65cm} K{1.65cm}}
\hline
\hline
\rule{0pt}{3ex}  
Planet & $M_p$ ($M_J$) & $X_{\mr{Na}}$\tnote{\textdagger} & $X_{\mr{K}}$\tnote{\textdagger} & $X_{\mr{H_2O}}$\tnote{\textdagger} & $X_{\mr{CH_4}}$\tnote{\textdagger} & $X_{\mr{NH_3}}$\tnote{\textdagger} & $X_{\mr{HCN}}$\tnote{\textdagger} & $X_{\mr{CO}}$\tnote{\textdagger} & $X_{\mr{CO_2}}$\tnote{\textdagger}\\ [1ex]
\hline
\rule{0pt}{3ex}
\\
 HAT-P-12b & $0.21^{+0.01}_{-0.01}$ & ${-8.75}^{+2.92}_{-2.16}$ & ${-3.24}^{+0.89}_{-4.25}$ & $\mathbf{ {-3.91}^{+1.01}_{-1.89}}$ & ${-9.14}^{+2.00}_{-1.82}$ & ${-8.98}^{+2.10}_{-1.94}$ & ${-8.43}^{+2.64}_{-2.28}$ & ${-5.65}^{+2.77}_{-4.02}$ & ${-6.16}^{+2.21}_{-3.39}$\\[1ex]
 
 \\
 
 WASP-39b & $0.28^{+0.03}_{-0.03}$ &${-3.86}^{+1.31}_{-1.36}$ & ${-4.22}^{+1.25}_{-1.12}$ & $\mathbf{ {-4.07}^{+0.72}_{-0.78}}$ & ${-5.65}^{+0.78}_{-1.02}$ & ${-8.65}^{+2.28}_{-2.16}$ & ${-8.24}^{+2.53}_{-2.43}$ & ${-7.02}^{+3.08}_{-3.22}$ & ${-4.31}^{+0.91}_{-1.04}$\\[1ex]
 
 \\
 
 WASP-6b & $0.50^{+0.02}_{-0.04}$ &${-9.18}^{+2.14}_{-1.82}$ & ${-5.53}^{+2.01}_{-1.85}$ & $\mathbf{ {-6.91}^{+1.83}_{-2.07}}$ & ${-9.15}^{+1.97}_{-1.83}$ & ${-8.62}^{+2.20}_{-2.17}$ & ${-8.60}^{+2.25}_{-2.18}$ & ${-7.77}^{+2.93}_{-2.69}$ & ${-10.08}^{+1.76}_{-1.29}$\\[1ex]
 
 \\
 
 HD 189733b & $1.14^{+0.03}_{-0.03}$ &${-7.77}^{+1.64}_{-0.87}$ & ${-8.87}^{+1.38}_{-1.52}$ & $\mathbf{ {-5.04}^{+0.46}_{-0.30}}$ & ${-9.18}^{+1.86}_{-1.84}$ & ${-9.09}^{+1.99}_{-1.91}$ & ${-8.13}^{+2.38}_{-2.55}$ & ${-6.93}^{+2.97}_{-3.33}$ & ${-9.02}^{+1.84}_{-1.96}$\\[1ex]
 
 \\
 
 HAT-P-1b & $0.53^{+0.02}_{-0.02}$ &${-8.44}^{+1.45}_{-2.12}$ & ${-9.10}^{+1.92}_{-1.84}$ & $\mathbf{ {-2.72}^{+0.42}_{-0.56}}$ & ${-8.39}^{+2.37}_{-2.29}$ & ${-7.85}^{+2.66}_{-2.64}$ & ${-7.65}^{+2.78}_{-2.79}$ & ${-7.29}^{+3.15}_{-3.05}$ & ${-8.93}^{+2.11}_{-1.96}$\\[1ex]
 
 \\
 
 HD 209458b & $0.69^{+0.02}_{-0.02}$ &${-4.92}^{+0.83}_{-0.57}$ & ${-6.46}^{+0.84}_{-0.64}$ & $\mathbf{ {-4.66}^{+0.39}_{-0.30}}$ & ${-8.60}^{+2.20}_{-2.22}$ & ${-8.56}^{+2.22}_{-2.28}$ & ${-8.66}^{+2.25}_{-2.21}$ & ${-8.37}^{+2.57}_{-2.38}$ & ${-9.80}^{+1.57}_{-1.45}$\\[1ex]
 
 \\
 
 WASP-31b & $0.48^{+0.03}_{-0.03}$ &${-7.66}^{+2.76}_{-2.68}$ & ${-3.43}^{+0.95}_{-1.62}$ & $\mathbf{ {-3.97}^{+1.01}_{-2.27}}$ & ${-8.30}^{+2.42}_{-2.34}$ & ${-5.01}^{+1.52}_{-4.45}$ & ${-7.92}^{+2.64}_{-2.59}$ & ${-7.32}^{+3.15}_{-3.01}$ & ${-9.20}^{+2.09}_{-1.83}$\\[1ex]
 
 \\
 
 WASP-17b & $0.49^{+0.03}_{-0.03}$ &${-8.71}^{+1.55}_{-1.76}$ & ${-9.77}^{+1.55}_{-1.38}$ & $\mathbf{ {-4.04}^{+0.91}_{-0.42}}$ & ${-9.26}^{+1.76}_{-1.73}$ & ${-8.33}^{+2.34}_{-2.32}$ & ${-8.70}^{+2.25}_{-2.10}$ & ${-5.41}^{+2.46}_{-4.13}$ & ${-7.04}^{+1.86}_{-2.77}$\\[1ex]
 
 \\
 
 WASP-19b & $1.11^{+0.04}_{-0.04}$ &${-7.51}^{+2.75}_{-2.80}$ & ${-7.10}^{+2.48}_{-2.63}$ & $\mathbf{ {-3.90}^{+0.95}_{-1.16}}$ & ${-8.76}^{+2.16}_{-2.05}$ & ${-8.44}^{+2.40}_{-2.23}$ & ${-8.03}^{+2.71}_{-2.48}$ & ${-6.45}^{+3.04}_{-3.48}$ & ${-7.03}^{+2.28}_{-2.97}$\\[1ex]
 
 \\
 
 WASP-12b & $1.40^{+0.10}_{-0.10}$ &${-4.11}^{+1.30}_{-1.75}$ & ${-8.88}^{+2.29}_{-2.00}$ & $\mathbf{ {-3.16}^{+0.66}_{-0.69}}$ & ${-8.84}^{+2.09}_{-1.99}$ & ${-8.47}^{+2.28}_{-2.19}$ & ${-8.17}^{+2.47}_{-2.41}$ & ${-7.18}^{+3.10}_{-3.00}$ & ${-8.34}^{+2.43}_{-2.32}$\\[1ex]
\\ 
\hline
\hline
\end{tabular}
\begin{tablenotes}
  \item[\textdagger] All values are in log$_{10}$($X_i$).
\end{tablenotes}
\end{threeparttable}
\end{sidewaystable*}

\begin{table*}
\parbox{11cm}{\caption{Retrieved cloud/haze properties of the hot Jupiter sample with the properties defined in equations (4-5).}
\label{tab:retvals2}}
\centering
\begin{tabular}{K{1.2cm} K{1.6cm} K{1.6cm} K{1.75cm} K{1.7cm} K{1.6cm}}
\hline
\hline
\rule{0pt}{3ex}  
Planet & $M_p$ ($M_J$) & log$_{10}$($a$)\tnote{\textdagger} & $\gamma$ & log$_{10}$($P_{\mr{cloud}}$)\tnote{\textdagger} & $\bar{\phi}$\\ [1ex]
\hline
\rule{0pt}{3ex}
\\
 HAT-P-12b & $0.21^{+0.01}_{-0.01}$ & ${4.70}^{+1.66}_{-1.74}$ & ${-9.04}^{+4.62}_{-7.78}$ & ${-0.78}^{+1.80}_{-2.03}$ & ${0.60}^{+0.24}_{-0.19}$\\[1ex]
 
 \\
 
 WASP-39b & $0.28^{+0.03}_{-0.03}$ &${3.83}^{+1.22}_{-1.10}$ & ${-13.27}^{+4.29}_{-4.24}$ & ${-1.09}^{+2.01}_{-2.53}$ & ${0.54}^{+0.24}_{-0.18}$\\[1ex]
 
 \\
 
 WASP-6b & $0.50^{+0.02}_{-0.04}$ &${4.92}^{+2.04}_{-1.41}$ & ${-6.06}^{+1.99}_{-2.63}$ & ${-0.10}^{+1.34}_{-1.38}$ & ${0.64}^{+0.17}_{0.15}$\\[1ex]
 
 \\
 
 HD 189733b & $1.14^{+0.03}_{-0.03}$ & ${3.64}^{+0.79}_{-1.01}$ & ${-7.75}^{+1.29}_{-1.43}$ & ${0.44}^{+1.03}_{-1.13}$ & ${0.67}^{+0.26}_{-0.12}$\\[1ex]
 
 \\
 
 HAT-P-1b & $0.53^{+0.02}_{-0.02}$ &${5.68}^{+1.52}_{-1.49}$ & ${-5.77}^{+2.49}_{-3.09}$ & ${-1.02}^{+1.93}_{-2.03}$ & ${0.52}^{+0.11}_{-0.10}$\\[1ex]
 
 \\
 
 HD 209458b & $0.69^{+0.02}_{-0.02}$ &${4.57}^{+0.58}_{-0.74}$ & ${-14.82}^{+4.79}_{-3.45}$ & ${-4.47}^{+0.52}_{-0.48}$ & ${0.52}^{+0.06}_{-0.07}$\\[1ex]
 
 \\
 
 WASP-31b & $0.48^{+0.03}_{-0.03}$ &${4.00}^{+0.95}_{-0.98}$ & ${-14.08}^{+4.07}_{-3.69}$ & ${-3.72}^{+0.66}_{-0.61}$ & ${0.76}^{+0.13}_{-0.15}$\\[1ex]
 
 \\
 
 WASP-17b & $0.49^{+0.03}_{-0.03}$ & ${2.33}^{+2.20}_{-2.53}$ & ${-10.55}^{+5.66}_{-5.84}$ & ${-1.60}^{+2.21}_{-2.56}$ & ${0.18}^{+0.26}_{-0.12}$\\[1ex]
 
 \\
 
 WASP-19b & $1.11^{+0.04}_{-0.04}$ &${3.95}^{+2.00}_{-2.22}$ & ${-11.52}^{+5.29}_{-5.11}$ & ${-0.56}^{+1.60}_{-1.97}$ & ${0.42}^{+0.27}_{-0.26}$\\[1ex]
 
 \\
 
 WASP-12b & $1.40^{+0.10}_{-0.10}$ & ${2.07}^{+2.64}_{-2.87}$ & ${-10.47}^{+7.11}_{-6.47}$ & ${-1.48}^{+2.14}_{-2.15}$ & ${0.33}^{+0.34}_{-0.20}$\\[1ex]
\\ 
\hline
\hline
\end{tabular}
\end{table*}

\begin{table*}
\parbox{10.5cm}{\caption{Retrieved temperature profile parameters of the hot Jupiter sample.}
\label{tab:retvals3}}
\centering
\begin{threeparttable}
\begin{tabular}{K{1.2cm} K{1.4cm} K{1.4cm} K{1.4cm} K{1.6cm} K{1.6cm} K{1.6cm} K{1.4cm}}
\hline
\hline
\rule{0pt}{3ex}  
Planet & $T_0$ & $\alpha_1$ & $\alpha_2$ & $P_1$\tnote{\textdagger} & $P_2$\tnote{\textdagger} & $P_3$\tnote{\textdagger} & $P_{\mr{ref}}$\tnote{\textdagger}\\ [1ex]
\hline
\rule{0pt}{3ex}
\\
 HAT-P-12b & ${456}^{+70}_{-40}$ & ${0.81}^{+0.13}_{-0.17}$ & ${0.63}^{+0.24}_{-0.28}$ & ${-1.54}^{+1.63}_{-1.74}$ & ${-3.97}^{+1.74}_{-1.34}$ & ${0.61}^{+0.94}_{-1.32}$ & ${-2.68}^{+0.84}_{-0.40}$ \\[1ex]
 
 \\
 
 WASP-39b & ${775}^{282}_{-166}$ & ${0.66}^{+0.22}_{-0.21}$ & ${0.59}^{+0.27}_{-0.26}$ & ${-1.50}^{+1.64}_{-1.77}$ & ${-3.99}^{+1.92}_{-1.35}$ & ${0.60}^{+0.95}_{-1.37}$ & ${-2.55}^{+0.52}_{-0.39}$\\[1ex]
 
 \\
 
 WASP-6b & ${1057}^{+198}_{-290}$ & ${0.57}^{+0.26}_{-0.22}$ & ${0.60}^{+0.26}_{-0.25}$ & ${-1.70}^{+1.61}_{-1.65}$ & ${-4.06}^{+1.79}_{-1.30}$ & ${0.49}^{+1.00}_{-1.33}$ & ${-3.55}^{+0.60}_{-0.74}$ \\[1ex]
 
 \\
 
 HD 189733b & ${1159}^{+146}_{-157}$ & ${0.76}^{+0.16}_{-0.22}$ & ${0.67}^{+0.22}_{-0.27}$ & ${-1.70}^{+1.75}_{-1.83}$ & ${-3.98}^{+1.85}_{-1.35}$ & ${0.48}^{+1.05}_{-1.42}$ & ${-2.07}^{+0.23}_{-0.43}$ \\[1ex]
 
 \\
 
 HAT-P-1b & ${1114}^{+251}_{-205}$ & ${0.65}^{+0.22}_{-0.22}$ & ${0.59}^{+0.26}_{-0.25}$ & ${-1.58}^{+1.61}_{-1.72}$ & ${-4.04}^{+1.83}_{-1.31}$ & ${0.57}^{+0.95}_{-1.30}$ & ${-3.63}^{+0.42}_{-0.41}$ \\[1ex]
 
 \\
 
 HD 209458b & ${949}^{+252}_{-109}$ & ${0.59}^{+0.27}_{-0.20}$ & ${0.35}^{+0.38}_{-0.17}$ & ${-1.85}^{+1.66}_{-1.28}$ & ${-4.22}^{+1.80}_{-1.22}$ & ${0.05}^{+1.29}_{-1.07}$ & ${-2.65}^{+0.39}_{-0.43}$ \\[1ex]
 
 \\
 
 WASP-31b & ${1043}^{+287}_{-172}$ & ${0.69}^{+0.21}_{-0.23}$ & ${0.61}^{+0.25}_{-0.26}$ & ${-1.64}^{+1.65}_{-1.75}$ & ${-4.04}^{+1.84}_{-1.32}$ & ${0.54}^{+0.98}_{-1.37}$ & ${-3.43}^{+0.61}_{-0.53}$ \\[1ex]
 
 \\
 
 WASP-17b & ${1147}^{+259}_{-305}$ & ${0.58}^{+0.26}_{-0.21}$ & ${0.58}^{+0.26}_{-0.24}$ & ${-1.52}^{+1.51}_{-1.68}$ & ${-3.99}^{+1.82}_{-1.33}$ & ${0.55}^{+0.95}_{-1.26}$ & ${-2.38}^{+0.32}_{-0.66}$ \\[1ex]
 
 \\
 
 WASP-19b & ${1386}^{+370}_{-337}$ & ${0.66}^{+0.22}_{-0.22}$ & ${0.61}^{+0.24}_{-0.24}$ & ${-1.69}^{+1.64}_{-1.66}$ & ${-3.99}^{+1.74}_{-1.33}$ & ${0.49}^{+0.98}_{-1.33}$ & ${-2.83}^{+0.89}_{-0.76}$\\[1ex]
 
 \\
 
 WASP-12b & ${990}^{+169}_{-122}$ & ${0.77}^{+0.15}_{-0.18}$ & ${0.66}^{+0.22}_{-0.26}$ & ${-1.55}^{+1.65}_{-1.71}$ & ${-3.92}^{+1.79}_{-1.35}$ & ${0.60}^{+0.93}_{-1.38}$ & ${0.16}^{+0.64}_{-0.63}$ \\[1ex]
\\ 
\hline
\hline
\end{tabular}
\begin{tablenotes}
  \item[\textdagger] All values are in log$_{10}$($P_i$[bar]).
\end{tablenotes}
\end{threeparttable}
\end{table*}

\bsp
\label{lastpage}
\end{document}